\documentclass [prc,aps,showpacs]{revtex4}
\usepackage{graphicx}
\begin{document}

\newcommand{\adag}{a^{\dag}}
\newcommand{\atil}{\tilde{a}}
\def\frp*1{${*1\over2}^+$}
\def\frm*1{${*1\over2}^-$}
\def\g{\noindent}
\def\mev{\hbox{\MeV}}
\def\kev{\hbox{\keV}}
\def\lambdabar{{\mathchar'26\mkern-9mu\lambda}}
\def\lambdabarrr{{^-\mkern-12mu\lambda} }

\title{Extensive statistical mechanics based on
nonadditive entropy: Canonical ensemble}

\author{A.S.~Parvan}

\affiliation{Bogoliubov Laboratory of Theoretical Physics, Joint
Institute for Nuclear Research, 141980 Dubna, Russia}
\affiliation{Institute of Applied Physics, Moldova Academy of
Sciences, MD-2028 Kishineu, Republic of Moldova}

\begin{abstract}
The original canonical ensemble formalism for the nonextensive
entropy thermostatistics is reconsidered. It is shown that the
unambiguous connection of the statistical mechanics with the
equilibrium thermodynamics is provided if the entropic parameter
$1/(q-1)$ is an extensive variable of the state. Based on a
particular example of the perfect gas, it is proved that the
Tsallis thermostatistics meets all the requirements of equilibrium
thermodynamics in the thermodynamic limit. In particular, the
entropy of the system is extensive and the temperature is
intensive. However, for finite systems both the Tsallis and
Boltzmann-Gibbs entropies are nonextensive. The equivalence of the
canonical and microcanonical ensembles of Tsallis thermostatistics
in the thermodynamic limit is established. The issue associated
with physical interpretation of the entropic variable is discussed
in detail.
\end{abstract}



\maketitle

\section{Introduction}
The main purpose of this Letter is to establish a clear way to
implement the equilibrium statistical mechanics based on the
nonadditive entropy deferent from the usual Boltzmann-Gibbs
statistical one. For the first time this concept was formulated by
Tsallis in~\cite{Tsal88}. It is very well known that the
conventional equilibrium statistical mechanics based on the
Boltzmann-Gibbs entropy meets all the requirements of the
equilibrium thermodynamics in the thermodynamic
limit~\cite{Balescu}. This is a necessary condition for
self-consistent definition of any equilibrium statistical
mechanics. In order to provide the connection of the statistical
mechanics with the thermodynamics, the statistical entropy is
usually used. From the mechanical and thermodynamical laws it
allows one to determine a unique phase distribution function, or a
statistical operator, which depends on two different sets of
variables: the first set specifies the dynamic state of the
microscopic system and the second one sets up the thermodynamic
state of the macroscopic system. According to the Liouville and
von Neumann equations, the equilibrium distribution function is a
constant of motion which is expressed only through the first
additive integrals of motion of the system. The ensemble averages
in the statistical mechanics correspond to the concrete functions
of state from the thermodynamics and depend only on the
macroscopic variables of state. The statistical entropy as a
function of the variables of state in the thermodynamic limit must
satisfy all properties of the thermodynamic entropy: concavity,
extensivity and so one. Note that the thermodynamic potentials of
the system are functions fixing the norm of the phase
distributions. It is known that the equilibrium thermodynamics is
the theory defined in the thermodynamic limit. Therefore, the
concept of the thermodynamic limit plays a crucial role in
comparing the equilibrium statistical mechanics with
thermodynamics. In this case, for thermodynamic systems the
boundary effects must be neglected and only the short-range
interaction forces can be taken into account. In the thermodynamic
limit the ensemble averages with the corresponding distribution
function should provide performance of the zero, first, second,
and third laws of thermodynamics, and the principle of additivity
that divide all variables into extensive and intensive ones.
Moreover, the fundamental equation of thermodynamics, the
Gibbs-Duhem relation, and the Euler theorem should be implemented.

In ref.~\cite{Tsal88}, the author tried to construct the
equilibrium statistical mechanics based on the nonextensive
statistical entropy. The problems arisen in the proof of the
thermodynamical laws resulted in the occurrence of the divers
variants of the Tsallis
thermostatistics~\cite{Tsal98,Martinez,Wang1}. In these
investigations not only the ensemble averages and the norm
equation for the distribution function but also the Lagrange
function were drastically redefined. Moreover, it was found that
any of these variants do not satisfy the zeroth law of
thermodynamics if the parameter $q$ must be a universal constant.
See ref.~\cite{Nauenberg} for subsequent discussions of the
critique of q-entropy for thermal statistics. To solve this
problem, in Abe et al.~\cite{Abe0} in the framework of the
canonical ensemble the physical temperature and pressure
satisfying the zeroth law of thermodynamics were introduced.
However, as it was shown in~\cite{Toral,Vives,Botet1,Parv1} such
transformations of the variables including the entropy lead to the
transition from the Tsallis thermostatistics to the extensive
statistical mechanics of Gibbs or R\'{e}nyi one. So such
consideration does not pertain to the perception of the zeroth law
of thermodynamics. In~\cite{Wang2}, a stronger assumption was
evolved to make use of the nonextensive microscopic Hamiltonian of
special kind like the nonextensive entropy form. In this case, the
Hamiltonian depends on the temperature of the system that entails
the changes of the thermodynamic relations and leads to the loss
of self-consistency of the statistical mechanics. Closely
connected with the problem of the zeroth law of thermodynamics is
another one: the principle of additivity in the thermodynamic
limit. In Abe~\cite{Abe1}, it was attempted to define the
thermodynamic limit on a particular example of the perfect gas in
the canonical ensemble. However, this limit was carried out
incorrectly. The correct definition of the thermodynamic limit for
the Tsallis thermostatistics in a particular case was given in
Botet et al.~\cite{Botet1,Botet2} and for the general case was
developed in~\cite{Parv2}. Note that an important criterion of
correctness of the thermodynamic limit is the equivalence of all
ensembles. Derivation of the fundamental equation of
thermodynamics for the Tsallis thermostatistics on the base of the
canonical distribution function was performed
in~\cite{Vives,Parv1}. In~\cite{Parv2}, it was proved that the
microcanonical ensemble of the Tsallis statistical mechanics
satisfies all requirements of the equilibrium thermodynamics if
the entropic index $1/(q-1)$ is the extensive variable of state of
the system. In the present Letter, we will show that similar
results are carried out also for the canonical ensemble.

The Letter is organized as follows. In the second section, the
canonical ensemble and the derivation of the thermodynamic
relations are given. In the third section, the performance of the
thermodynamic principles in the thermodynamic limit on the example
of the perfect gas is proved.

\section{Canonical Ensemble}
Let us consider the equilibrium statistical ensemble of the
classical dynamical systems of $N$ particles at the constant
temperature $T$, the volume $V$, and the thermodynamic coordinate
$z$ in a thermal contact with a heat bath. The system interacts
weakly with its surroundings and only the energy can be
transferred in and out of it. In order to determine the
equilibrium distribution function, we consider the Tsallis
equilibrium statistical entropy which is a function of the
parameter $q$ and a functional of the probing phase distribution
function $\varrho(x,p)$:
\begin{equation}\label{1}
    S=-k\int \frac{\varrho-\varrho^{q}}{1-q}\  d\Gamma ,
\end{equation}
where $d\Gamma =dxdp$ is an infinitesimal element of phase space,
$k$ is the Boltzmann constant and $q\in\mathbf{R}$ is the real
parameter taking values $0<q<\infty$. The phase distribution
function is normalized to unity:
\begin{equation}\label{2}
    \int \varrho \ d\Gamma =1.
\end{equation}
In the classical statistical mechanics the expectation value of
the Hamiltonian can be written as
\begin{equation}\label{3}
    \langle H\rangle =\int \varrho H \ d\Gamma.
\end{equation}
The phase distribution function depends on the first additive
constants of motion of the system. Nevertheless, the mechanical
laws are not sufficient to determine it unambiguously. For this
reason additional postulates of the equilibrium thermodynamics are
required. To express the equilibrium phase distribution function
from the macroscopic variables of state, we consider the
thermodynamic method explored in~\cite{Parv2}. In the state of
thermal equilibrium the macroscopic system is characterized by the
fundamental equation of thermodynamics
\begin{equation}\label{4}
    T dS_{\mathrm{th}}=dE+p dV+X dz-\mu dN,
\end{equation}
where $S_{\mathrm{th}}(T,V,z,N)$ is the thermodynamic entropy, $z$
and $V$ are the "thermodynamic coordinates"; $X$ and $p$ are the
associated "forces"; $\mu$ is the chemical potential and $E$ is
the thermodynamic energy of the system. In the canonical ensemble
the fundamental equation of thermodynamics at the fixed values of
$T,V,z,N$ can be rewritten as
\begin{equation}\label{5}
    (T dS_{\mathrm{th}}-dE)_{T,V,z,N}=0.
\end{equation}
Then, to express the phase distribution function $\varrho(x,p)$
through the variables of state $(T,V,z,N)$, let us replace in
Eq.~(\ref{5}) the equilibrium thermodynamic entropy
$S_{\mathrm{th}}$ and energy $E$ of the macroscopic system with
the statistical ones (\ref{1}) and (\ref{3}). Then, one finds
\begin{eqnarray}\label{6}
    T\frac{\partial S}{\partial q}dq +
    \int  d\Gamma \left\{ \left[T \frac{\delta S}{\delta \varrho}-
    \frac{\delta \langle H\rangle}{\delta \varrho}\right] d\varrho -
     \frac{\delta\langle H\rangle}{\delta H}\ dH \right\} = 0,
\end{eqnarray}
where the symbol $d$ before the functions $H$, $\varrho$ and $q$
is the total differential in variables $(T,V,z,N)$. The
statistical parameters $q$ and $\varrho$ must be expressed through
the variables of state of the system to provide the unambiguous
conformity between statistical and thermodynamic entropies. Since
$dH=0$, $dq=0$ and $d\varrho =0$, we obtain
\begin{equation}\label{8}
    T\frac{\delta S}{\delta \varrho}-
    \frac{\delta \langle H\rangle}{\delta \varrho}=\alpha,
\end{equation}
where $\alpha$ is a certain constant. For the microcanonical
ensemble it was stated in ref.~\cite{Parv2} that
\begin{equation} \label{9}
  \frac{1} {q-1}= z,
\end{equation}
where the parameter $z$ takes the values $-\infty < z < -1$ for
$0<q < 1$ and $0<z<\infty$ for $1< q<\infty$ and in the limiting
case for $q=1$, we have $z =\pm\infty$. Substituting Eq.~(\ref{1})
into (\ref{8}) and using again equation (\ref{1}) to eliminate the
parameter $\alpha$ we arrive at the following expression for the
equilibrium phase distribution function:
\begin{equation}\label{10}
\varrho=\left[1+\frac{1}{z+1}\frac{\Lambda-H}{kT}\right]^{z},
\end{equation}
where $\Lambda=E-\frac{z+1}{z}\ TS$ and it is determined from the
normalization condition (\ref{2})
\begin{equation}\label{11}
    \int \left[1+\frac{1}{z+1}\frac{\Lambda-H}{kT}\right]^{z}
    d\Gamma=1.
\end{equation}
Thus, $\Lambda$ is a function of the variables of state, $\Lambda
=\Lambda(T,V,z,N)$. For the Jaynes principle derivation of the
phase distribution function (\ref{10}), see Appendix A. The
expectation value $\langle A \rangle$ of the dynamical variable
$A(x,p)$ can be defined as follows (cf.~(\ref{3})):
\begin{equation}\label{12}
\langle A \rangle =\int A
\left[1+\frac{1}{z+1}\frac{\Lambda-H}{kT}\right]^{z} d\Gamma.
\end{equation}
Using Eqs.~(\ref{2}), (\ref{3}) and (\ref{10}), we can write the
entropy (\ref{1}) as
\begin{equation}\label{13}
    S=\frac{z}{z+1}\ \frac{\langle H\rangle
    -\Lambda}{T}.
\end{equation}
In the canonical ensemble it is convenient to introduce the  free
energy as a thermodynamic potential:
\begin{equation}\label{14}
    F\equiv E-TS=\frac{\langle H\rangle+z
    \Lambda}{z+1} , \;\;\;\;\;\;\;\;\; E=\langle H\rangle.
\end{equation}
This is the Legendre transform of the energy with respect to the
entropy of the system with $\partial E/\partial S=T$.

At this point we have obtained a rigorous derivation of classical
statistical mechanics. The quantum statistical mechanics is
constructed in analogy with the classical one. In this respect,
all classical notions are replaced by the quantum mechanical ones.
A dynamic state of the quantum system is defined by a vector of
state $|\Psi(t)\rangle$, which is an element of the abstract
Hilbert space $\mathcal{E}_{\mathrm{H}}$. The dynamic variables
are represented by linear hermitian operators $A$ acting on the
elements of the Hilbert space. In particular, the Hamiltonian $H$
is the linear hermitian operator acting on the vectors of state
$|\Psi(t)\rangle$. In the quantum statistical mechanics the mixed
states are considered. A macrostate thus appears as a set of
possible microstates, which are set up by state vectors
$|\Psi_{r}(t)\rangle$, $r=1,2,\ldots$, each with its own
probability $w_{r}$ for its occurrence, which are eigenvalues of
the statistical operator, $\varrho(t)|\Psi_{r}(t)\rangle = w_{r}
|\Psi_{r}(t)\rangle$. The integral over phase space of the
classical functions is replaced by the trace of the corresponding
quantum operators.

Let us show the connection between the statistical mechanics and
equilibrium thermodynamics. In this respect, we derive the
thermodynamic relations for the canonical ensemble from the
general point of view. Let us consider now the classical case.
Applying the total differential operator with respect to the
ensemble variables $(T,V,z,N)$ on the entropy (\ref{1}) and the
norm equation (\ref{2}), and using Eq.~(\ref{10}) one finds
\begin{equation}\label{1b}
   T dS= Xdz+ \int d\varrho H d\Gamma,
\end{equation}
where
\begin{equation}\label{1bb}
    X=kT\int \varrho [1-\varrho^{1/z}(1-\ln\varrho^{1/z})]
    d\Gamma.
\end{equation}
The repeated application of this differential operator to
Eq.~(\ref{3}) and substitution of the results into Eq.~(\ref{1b})
leads to the formula
\begin{equation}\label{2b}
    T dS= d\langle H\rangle- \int \varrho dH d\Gamma+Xdz.
\end{equation}
By virtue of the parametrical dependence of the Hamilton function
$H$ on the variables $V$ and $N$ we have the fundamental equation
of thermodynamics~\cite{Vives,Parv1}
\begin{equation}\label{3b}
    T dS= d\langle H\rangle + pdV +X dz -\mu dN,
\end{equation}
where
\begin{eqnarray}\label{4b}
  p &=& \int \varrho
 \left(-\frac{\partial H}{\partial V}
 \right)_{T,z,N} d\Gamma =\left\langle -\frac{\partial H}{\partial V}
 \right\rangle , \\ \label{5b}
  \mu &=& \int \varrho
 \left(\frac{\partial H}{\partial N}
 \right)_{T,V,z} d\Gamma =\left\langle \frac{\partial H}{\partial N}
 \right\rangle .
\end{eqnarray}
Here, the property of the Hamilton function $(\partial H/\partial
T)_{V,z,N}=(\partial H/\partial z)_{T,V,N}=0$ is used. So we have
proved that the statistical entropy (\ref{1}) with the phase
distribution function (\ref{10}) completely satisfies  the
fundamental equation of thermodynamics (\ref{4}) which was a
starting point of our derivations concerning the distribution
function. In fact, the differential form (\ref{3b}) is sufficient
to prove the connection of the thermodynamics and statistical
mechanics.

The thermodynamic potential of the canonical ensemble $(T,V,z,N)$
is the Helmholtz free energy (\ref{14}). Hence, the differential
of $F$ which follows from Eq.~(\ref{3b}) can be written as
\begin{equation}\label{6b}
    dF= -S dT - p dV- X dz + \mu dN,
\end{equation}
and the thermodynamic relations are
\begin{eqnarray}\label{7b}
  S &=& -\left(\frac{\partial F}{\partial T}\right)_{V,z,N}, \;\;\;\;\;\;\;\;
  p = -\left(\frac{\partial F}{\partial V}\right)_{T,z,N}, \\ \label{8b}
  X &=& -\left(\frac{\partial F}{\partial z}\right)_{T,V,N},
  \;\;\;\;\;\;\;\;\; \mu = \left(\frac{\partial F}{\partial
  N}\right)_{T,V,z}.
\end{eqnarray}
The free energy as a thermodynamic potential can easily be
calculated in the framework of the canonical ensemble and it is used
in order to obtain the functions of the state.

The fundamental equation of thermodynamics (\ref{3b}) provides the
first and second principles of thermodynamics
\begin{equation}\label{9b}
    \delta Q=TdS, \;\;\;\;\;\;\;\;\; \delta Q= d\langle H\rangle + pdV +X dz -\mu
    dN,
\end{equation}
where $\delta Q$ is a heat transfer by the system to the
environment during a quasistatic transition of the system from one
equilibrium state to a nearby one.

Let us now find the important quantity, the heat capacity, which
is defined from the general rule, $\delta Q =C dT$. According to
the first and second laws of thermodynamics, in the canonical
ensemble the heat capacity at the fixed values of $T,V,z,N$ can be
written as
\begin{eqnarray}\label{10b} \nonumber
    C_{VzN}(T,V,z,N)&=&\left(\frac{\delta Q}{dT}\right)_{V,z,N}=
    T\left(\frac{\partial S}{\partial T}\right)_{V,z,N}= \\
    &=&\left(\frac{\partial \langle H\rangle}{\partial T}\right)_{V,z,N}=
    -T\left(\frac{\partial^{2} F}{\partial T^{2}}\right)_{V,z,N}.
\end{eqnarray}
Note that all thermodynamic relations of classical statistical
mechanics that we have just given are also preserved in the case
of formalism starting from quantum mechanics.

It is important to note that the first formalism of the Tsallis
statistics~\cite{Tsal88} was rejected in the literature. For
details, see~\cite{Tsal98}. In the present study, we reconsider
this initial formalism and show that the Tsallis phase
distribution function was not accurately determined. It is claimed
in~\cite{Tsal88} that the dependent function of the state
$\beta(T,V,z,N)$, or the Lagrange multiplier, is an independent
parameter of the distribution function. However, we show that the
consistent derivation of the phase distribution function in terms
of the variables of state $(T,V,z,N)$, based on the Jaynes
principle with the Lagrange function introduced in~\cite{Tsal88},
leads exactly to our Eqs.(\ref{10}) and (\ref{11}). For details,
see Appendix A. Note that in the limit, $z\to\pm\infty$, all
expressions given above resulted in ones of the conventional Gibbs
thermostatistics.

\section{The thermodynamic limit. The Perfect Gas}
The statistical mechanics must satisfy the requirements of the
equilibrium thermodynamics in the thermodynamic limit when the
number of particles of the system considered is very large, and
the relative magnitude of surface effects becomes negligible. For
thermodynamic systems, which are homogeneous at a macroscopic
scale, the principle of additivity is valid: the various
quantities of interest can be classified into either extensive or
intensive ones under a division of the system into macroscopic
parts. The extensive quantity, for instance, the entropy
considered as a function of the extensive variables $V,z$ and $N$
is homogeneous of degree 1:
\begin{equation}\label{11b}
S(T,\lambda V,\lambda z,\lambda N)=\lambda S(T,V,z,N),
\end{equation}
where $\lambda$ is a certain constant. However, the intensive
quantities, like the pressure $p$, the chemical potential $\mu$,
and $X$, are the homogeneous functions of degree zero:
\begin{equation}\label{12b}
\mu(T,\lambda V,\lambda z,\lambda N)=\mu(T,V,z,N).
\end{equation}
Note that the temperature $T$ is an intensive variable of state
remaining invariant under such a subdivision of the system. This
invariance property of the temperature guarantees the fulfillment
of the zeroth law of thermodynamics. Here the thermodynamic limit
denotes the limiting statistical procedure $N\rightarrow\infty$,
$v=V/N =\mathrm{const}$, $\tilde{z}=z/N =\mathrm{const}$ with
keeping the main asymptotic on $N$. It is meant to make an
expansion of the functions of the state in powers of the small
parameter $1/N$ $(N\gg 1)$ with large finite values of the
variables $V,z$. Then the extensive variables $\mathcal{A}$ can be
written $(\alpha> 0)$ as
\begin{equation} \label{13b}
 \mathcal{A}(T,V,z,N)\left|_{N\rightarrow\infty,
      v,\tilde{z} = \mathrm{const} } \right. = N[a(T,v,\tilde{z}) +
      O(N^{-\alpha})]
      \stackrel{as} = N a(T,v,\tilde{z}),
\end{equation}
whereas the intensive variables $\phi$ take the following form:
\begin{equation} \label{14b}
\phi(T,V,z,N)\left|_{N\rightarrow\infty,
      v,\tilde{z} = \mathrm{const}} \right. = \phi(T,v,\tilde{z}) + O(N^{-\alpha})
      \stackrel{as} = \phi(T,v,\tilde{z}),
\end{equation}
where $v$ is the specific volume, $\tilde{z}$ is the specific $z$,
and $a =\mathcal{A} /N$ is the specific $\mathcal{A}$. It is
important to note that the limiting statistical procedures
$(V\to\infty, z/V=\mathrm{const}, N/V=\mathrm{const})$ and
$(z\to\pm\infty, V/z=\mathrm{const},N/z=\mathrm{const})$ are
equivalent with the limit, $N\to\infty$, given above. Note that in
Abe~\cite{Abe1}, the thermodynamic limit for the Tsallis
statistics is not correct because the limits $N\to\infty$ and
$|z|\to\infty$ are not coordinated among themselves. Note that
after applying the thermodynamic limit to the functions of state
the Boltzmann-Gibbs limit, $\tilde{z}\to\pm\infty$, is provided by
expansion of these functions in powers of the small parameter
$1/\tilde{z}$ holding only the zero term of the power expansion.

In the canonical ensemble, how to prove from the general point of
view the principle of additivity (see Eqs.~(\ref{11b}) and
(\ref{12b})) and the zeroth law is not obvious. Therefore, we will
illustrate explicitly the implementation of these principles on
the foregoing example of the nonrelativistic ideal gas. In the
framework of the nonrelativistic ideal gas of $N$ identical
particles in the canonical ensemble the functions of state can be
explicitly expressed from the variables of state. Hence, the
thermodynamic properties of the Tsallis statistics can thoroughly
be investigated. In order to evaluate the expectation values of
the dynamical variables, we use the method based on the integral
representation of the Euler gamma function~\cite{Prato,Parv1}. Let
us investigate the thermodynamic properties of the nonrelativistic
perfect gas of $N$ identical particles in the thermodynamic limit
($N\to \infty$, $v=\mathrm{const}$, $\tilde{z}=\mathrm{const}$).
The exact relations for it can be found in Appendix B. In the
thermodynamic limit one easily confirms that the canonical
partition function of the ideal gas for Gibbs statistics is
simply~\cite{Huang}
\begin{equation}\label{12d}
Z_{G}^{1/N}(T,V,N)=(gve)
\left(\frac{mkT}{2\pi\hbar^{2}}\right)^{3/2}\equiv
\tilde{Z}_{G}(T,v),
\end{equation}
where $\tilde{Z}_{G}$ is the one-particle partition function, $m$
is the particle mass and $g$ is the spin degeneracy factor. Then
in the thermodynamic limit and in the limit of Boltzmann-Gibbs
statistics, $\tilde{z}\to\pm\infty$, Eqs.~(\ref{1d}) and
(\ref{2d}) can be written as
\begin{eqnarray}\label{13d}
    B(T,V,z,N) &=&
    \left(\tilde{Z}_{G}e^{3/2}\right)^{-\frac{1}{\tilde{z}+\frac{3}{2}}}
    \ \left[1+\frac{3}{2\tilde{z}}\right]=B(T,v,\tilde{z}), \\
 \label{14d}
 \left. B \right|_{\tilde{z}\to\pm\infty} &=&
1-\frac{1}{\tilde{z}}\ln\tilde{Z}_{G}+O((1/\tilde{z})^{2}),
\end{eqnarray}
where $\Lambda/N=kT\tilde{z}(B-1)$ and $\tilde{z}<-3/2$ or
$\tilde{z}>0$ with respect to the conditions of the integration
method used. Note that the function of state $B$ is intensive and
$\Lambda$ is extensive. The energy (\ref{3d}) takes the following
form:
\begin{equation}\label{15d}
\frac{E(T,V,z,N)}{N}=\frac{3}{2}kT \
\frac{B}{1+\frac{3}{2\tilde{z}}}=\varepsilon(T,v,\tilde{z}),
\end{equation}
where $\varepsilon$ is the specific energy depending only on
intensive variables and it is intensive. Consequently, the energy
$E$ is extensive (cf. Eqs.(\ref{13b}) and (\ref{14b})). Note that
in the Boltzmann-Gibbs limit the energy (\ref{15d}) is reduced to
$E|_{\tilde{z}\to \pm\infty}=(3/2)kT N$. The free energy
(\ref{4d}) is found to be
\begin{equation}\label{16d}
    \frac{F(T,V,z,N)}{N}=kT \tilde{z}
    (B-1)=f(T,v,\tilde{z}),
\end{equation}
where $f$ is the specific free energy which is intensive and
$F=\Lambda$. In the Boltzmann-Gibbs limit the free energy
(\ref{16d}) takes the usual form $\left. F\right|_{\tilde{z}\to
\pm\infty}=-kT\ln Z_{G}$. Then in the thermodynamic limit and in
the limit of Boltzmann-Gibbs statistics the entropy (\ref{5d}) can
be written as
\begin{eqnarray}\label{17d}
    \frac{S(T,V,z,N)}{N}&=& k\tilde{z}
    \left[1-\frac{B}{1+\frac{3}{2\tilde{z}}}\right]=s(T,v,\tilde{z}),
    \\ \label{18d}
    \;\;\; \left. S \right|_{\tilde{z}\to
    \pm\infty}&=& \frac{3}{2}kN+k\ln Z_{G} =S_{G},
\end{eqnarray}
where $s$ is the specific entropy and $S_{G}$ is the entropy of
Gibbs statistics. The function $s$, however, is an intensive
function of $(T,v,\tilde{z})$. Thus, the Tsallis entropy
(\ref{17d}) in the thermodynamic limit is extensive. It is
important to note that the Tsallis entropy (\ref{5d}) for the
finite values of $N$ and $z$ does not satisfy the homogeneous
condition (\ref{11b}), for instance, in the case of $\lambda=1/N$,
and it is nonextensive. However, the Gibbs entropy, $S_{G}$, is
also nonextensive for the nonrelativistic perfect gas of $N$
identical particles in the canonical ensemble, because the
canonical partition function $Z_{G}\sim (N!)^{-1}$. See Appendix
B. In the thermodynamic limit the Gibbs entropy (\ref{18d}) of the
perfect gas in contrast with its exact value is extensive,
because Eq.~(\ref{12d}) has been implemented. Note that the values
of the temperature and the specific volume are restricted, $T\geq
T_{0}$ and $v\geq v_{0}$, by the physical conditions $S\geq 0$ and
$S_{G}\geq 0$, where the parameters $(T_{0},v_{0})$ are determined
from the following equation $\tilde{Z}_{G}(T_{0},v_{0})e^{3/2}=1$.

Let us deduce the pressure $p$, the chemical potential $\mu$, and
the variable $X$ in the thermodynamic limit. The pressure
(\ref{6d}) takes the form
\begin{equation}\label{19d}
    p(T,V,z,N)=\frac{kT}{v}\ \frac{B}{1+\frac{3}{2\tilde{z}}} =
    p(T,v,\tilde{z}),
\end{equation}
where $p$ is the intensive function and in the Gibbs limit we have
$\left. p\right|_{\tilde{z}\to \pm\infty}=kT/v$. The function of
state (\ref{7d}) can be written as
\begin{equation}\label{20d}
    X(T,V,z,N)=kT\left[1-\frac{B}{1+\frac{3}{2\tilde{z}}}\
    \left(1-\ln\frac{B}{1+\frac{3}{2\tilde{z}}}\right)\right]=X(T,v,\tilde{z})
\end{equation}
and in the Gibbs limit it is reduced to $\left.
X\right|_{\tilde{z}\to \pm\infty}=0$. The chemical potential
(\ref{8d}) is now
\begin{equation}\label{21d}
    \mu(T,V,z,N)=kT \frac{B}{1+\frac{3}{2\tilde{z}}}\
    \left(\frac{5}{2}+\tilde{z}\ln\frac{B}{1+\frac{3}{2\tilde{z}}}\right)=
    \mu(T,v,\tilde{z}),
\end{equation}
where $\mu$ is an intensive one and in the Gibbs limit we find
that $\left. \mu\right|_{\tilde{z}\to \pm\infty}=kT(1-\ln
\tilde{Z_{G}})$. Then, Eqs.~(\ref{15d}), (\ref{17d}), and
(\ref{19d})-(\ref{21d}) in the thermodynamic limit yield the Euler
theorem
\begin{equation}\label{22d}
TS = E+pV+Xz-\mu N.
\end{equation}
This equation enables us to interpret $X$ as the "force"
observable of the system conjugate to the "position" variable
$z$. Moreover, Eq.~(\ref{22d}) allows us to write
\begin{equation}\label{23d}
  F = E-TS=-pV-Xz+\mu N.
\end{equation}
Its differential leads to the Gibbs-Duhem relation:
\begin{equation}\label{24d}
    SdT=Vdp+zdX-Nd\mu,
\end{equation}
which shows that the variables $T,p,X$ and $\mu$ are not
independent. The heat capacity of the perfect gas of $N$ identical
particles is deduced from Eq.~(\ref{10b}). Using Eq.~(\ref{15d})
we obtain
\begin{equation}\label{25d}
C_{v\tilde{z}}(T,v,\tilde{z})=\frac{3}{2}kN
\frac{B}{(1+\frac{1}{\tilde{z}}\frac{3}{2})^{2}}.
\end{equation}
In the Boltzman-Gibbs limit the heat capacity takes its usual form
$\left. C_{v\tilde{z}}\right|_{\tilde{z}\to \pm\infty}=(3/2)kN$.

To better understand the thermodynamic properties of the perfect
gas, it is necessary to study the equilibrium distribution of the
momenta $\vec{p_{i}}$ of the gas particles. The $N$-particle
distribution function is defined in Appendix~B. In the
thermodynamic limit the single-particle distribution function
(\ref{11d}) takes the form
\begin{equation}\label{26d}
    f(\vec{p})=\left(\frac{1}{2\pi m_{eff}kT}\right)^{3/2} \
    e^{-\frac{\vec{p}^{2}}{2m_{eff}kT}},
\end{equation}
where $m_{eff}$ is the effective particle mass written as
\begin{equation}\label{27d}
    m_{eff}=m \ \frac{B}{1+\frac{1}{\tilde{z}}\frac{3}{2}}=
    \left\{m\left[
    gv\left(\frac{kTe^{5/3}}{2\pi\hbar^{2}}\right)^{3/2}
           \right]^{-1/\tilde{z}}
    \right\}^{1/(1+\frac{1}{\tilde{z}}\frac{3}{2})}.
\end{equation}
In the Boltzmann-Gibbs limit, $\left. m_{eff}\right|_{\tilde{z}\to
\pm\infty}=m$, and the single-particle distribution function is
reduced to the Maxwell-Boltzmann distribution $\left.
f(\vec{p})\right|_{\tilde{z}\to \pm\infty} =(2\pi mkT)^{-3/2}
\exp(-\vec{p}^{2}/2mkT)$. It is clearly seen that the effective
mass (\ref{27d}) is $m_{eff}\geq m$ for $\tilde{z}<-3/2$ and
$m_{eff}\leq m$ for $\tilde{z}>0$. At $\tilde{z}\to -3/2$ we have
$m_{eff}\to\infty$. Let us investigate the single-particle
averages. The particle mean kinetic energy  with the
single-particle distribution function (\ref{26d}) can be written
as
\begin{equation}\label{28d}
    \left\langle \frac{\vec{p}^{2}}{2m}\right\rangle  = \frac{3}{2} kT
    \frac{m_{eff}}{m}=\varepsilon.
\end{equation}
So the average kinetic energy is equivalent with the specific
energy per particle (\ref{15d}). The average momentum of the
particle and the highest probability momentum of the distribution
$f(p)$ can be written as
\begin{equation}\label{29d}
    \langle p\rangle =\sqrt{\frac{8m_{eff}kT}{\pi}},
    \;\;\;\;\;\;\;  p_{hp}=\sqrt{2m_{eff}kT},
\end{equation}
where $p=|\vec{p}|$. Their ratio is $\langle
p\rangle/p_{hp}=2/\sqrt{\pi}\approx 1.13$ as in the Gibbs
statistics for which the average momentum and the highest
probability momentum are $\langle p\rangle_{G}=(8mkT/\pi)^{1/2}$
and $(p_{hp})_{G}=(2mkT)^{1/2}$. The variance and the relative
statistical fluctuations for the distribution (\ref{26d}) are
\begin{eqnarray}\label{30d}
    \langle (\triangle p)^{2}\rangle &=& \langle p^{2}\rangle -\langle
    p\rangle^{2}=kT m_{eff} \left(3-\frac{8}{\pi}\right), \\
    \label{31d}
    \delta_{p}&=& \frac{\sqrt{\langle (\triangle p)^{2}\rangle}}{\langle
    p\rangle}= \sqrt{\frac{3\pi}{8}-1}
\end{eqnarray}
or $ \delta_{p}\approx 0.424$. So from the investigation of the
distribution function (\ref{26d}) and its averages we arrive at
the conclusion that in the Tsallis statistics the mean kinetic
energy of the particles and the momentum are larger than their
values in the Gibbs statistics, $\varepsilon \geq \varepsilon_{G}$
and $\langle p\rangle\geq \langle p\rangle_{G}$, for
$\tilde{z}<-3/2$ and smaller, $\varepsilon \leq \varepsilon_{G}$
and $\langle p\rangle\leq \langle p\rangle_{G}$, for
$\tilde{z}>0$. Now we can give a physical interpretation for the
variable of state $\tilde{z}$ in the framework of the Tsallis
thermostatistics if we consider the system of noninteracting
particles as the system of the interacting quasiparticles with the
effective mass $m_{eff}$. Then the total energy of the
quasiparticle, $\varepsilon$, is equal to the sum of the mean
kinetic energy and the effective interaction energy $\triangle
\varepsilon$ which can be written as
\begin{equation}\label{32d}
\left\langle \frac{\vec{p}^{2}}{2m_{eff}}\right\rangle =
\frac{3}{2} kT, \;\;\;\;\;\;\; \triangle \varepsilon = \frac{3}{2}
kT\left[\frac{m_{eff}}{m}-1\right].
\end{equation}
So the effective interaction energy is positive
$\triangle\varepsilon>0$, and the forces are repulsive for
$\tilde{z}<-3/2$ and they are attractive $\triangle \varepsilon<0$
for $\tilde{z}>0$. Note that from Eq.~(\ref{32d}) follows the
physical interpretation for the temperature of the system  $T$ as
the average kinetic energy of the quasiparticles with the
effective mass $m_{eff}$.

\begin{figure}
  \includegraphics[width=15cm]{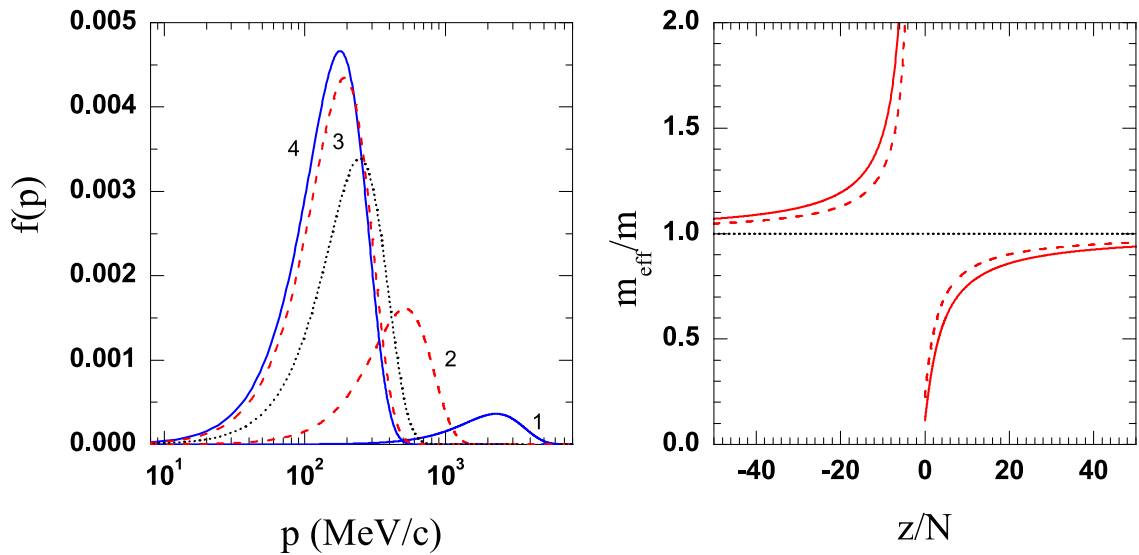}\\
  \caption{The dependence of the single-particle distribution function on
  the momentum $p$ (left) for the classical perfect gas of quarks with
  two flavor  and three color degrees of freedom in the non-relativistic approximation for the
different values of $\tilde{z}=-2$, $-3$, $3$ and $2$
  (the curves $1,2,3$ and $4$, respectively) at the temperature $T=100$ MeV
  and the specific volume $v=0.25/\rho_{0}$. The ratio of the effective quark mass to the
   constituent one as a function of the specific $\tilde{z}$ (right) for the values of $T=100$
    MeV (dashed line) and $T=200$ MeV (solid one).  The dotted lines correspond to the conventional
     Boltzmann-Gibbs thermostatistics. }\label{fg1}
\end{figure}

Fig.~1 shows the dependence of the single-particle distribution
function (left panel) on the momentum $p$ for the perfect gas of
free quarks in the nonrelativistic approximation. The calculations
are carried out for the system of quarks with two flavor and three
color degrees of freedom, and the constituent mass $m=300$ MeV at
temperature $T=100$ MeV and the specific volume $v=0.25/\rho_{0}$,
where $\rho_{0}=0.168 fm^{-3}$. Note that the quarks at short
distances are treated as almost free pointlike noninteracting
particles because of the property of asymptotic freedom. It is
remarkable that the single-particle distribution function at small
values of the variable $|\tilde{z}|$ considerably differs from the
limiting Maxwell-Boltzmann distribution. Such a behavior has
really been caused by the sharp changes of the effective mass of
quasiparticles in the dependence on $\tilde{z}$. This dependence
can be seen even better in the right panel of Fig.~1 which shows
the ratio of the effective mass to the constituent one vs. the
variable $\tilde{z}$ for two values of the temperature $T$.
Figure~1 clearly shows that the variable $\tilde{z}$  is the order
parameter and the system is physically unstable in the region
$-3/2<\tilde{z}<0$. For the microcanonical ensemble similar
results were obtained in~\cite{Parv2}.

Our exploration of the perfect gas has brought the following
points to the foreground: for the Tsallis statistics in the
thermodynamic limit the principle of additivity and the zeroth law
of thermodynamics are valid. The entropy (\ref{17d}) is a
homogeneous function of the first order, it is an extensive
variable satisfying the relation (\ref{11b}) with $\lambda=1/N$
provided that the temperature $T$ must be the intensive variable
of state. It should, however, be noted that in the thermodynamic
limit the equivalence of the canonical and microcanonical
ensembles is implemented. Giving  constants of motion exactly or
only as averages leads to the same results. For instance, if we
express the temperature $T$ through the variables $(E,V,z,N)$ from
Eq.~(\ref{15d}) as
\begin{equation}\label{33d}
    kT=\frac{2}{3} \varepsilon w^{1/\tilde{z}}, \;\;\;\;\;\;
    w=gv \left(\frac{m\varepsilon e^{5/3}}{3\pi \hbar^{2}}\right)^{3/2}
\end{equation}
and substitute it into Eqs.~(\ref{17d}) and
(\ref{19d})-(\ref{21d}), then we obtain the results of the
microcanonical ensemble derived in ref.~\cite{Parv2}. It is
important to note that in terms of the $z$ variable and our
thermodynamic limit the results for the perfect gas in the
canonical ensemble of Abe et al.~\cite{Abe1,Abe2} are the same as
here.

\section{Conclusions}
In this Letter, the canonical ensemble of the nonextensive
thermostatistics introduced by C.~Tsallis has been reconsidered.
It is shown that the equilibrium statistical mechanics based on
the nonadditive statistical entropy completely satisfies all
requirements of the equilibrium thermodynamics in the
thermodynamic limit. The unique non-Gibbs phase distribution
function corresponding to the Tsallis entropy is obtained from the
constraints imposed by the equilibrium thermodynamics laws. The
microscopic foundation of the equilibrium statistical mechanics
proceeds on the Gibbs idea of the statistical ensembles for the
quantum and classical mechanics. The phase distribution function
and the statistical operator depend only on the first additive
constants of motion of the system. Here they were derived within a
formalism based on the fundamental equation of thermodynamics and
statistical definition of the functions of state. It allows us to
avoid introduction of the controversial Lagrange multiplies.
Nevertheless, it is shown that the distribution function derived
from the Jaynes principle exactly coincides with ours if the
Lagrange parameters are expressed through a set of independent
variables of state of the system. The unambiguous connection of
the statistical mechanics with thermodynamics is established. The
equilibrium distribution function satisfies the fundamental
equation of thermodynamics, the first and the second principles
preserving the Legendre transformation. All thermodynamic
relations relative to the Helmholtz free energy for the
thermodynamic system in the thermostat are carried out. The heat
capacity of the system was derived from the first and the second
laws of thermodynamics. Note that in the fundamental equation of
thermodynamics the new term related to the work of the conjugate
force $X$ at changes of the variable of state $z$ appeared. In the
limit $z\to\pm\infty$ the conventional Gibbs statistics is
recovered.

It is well known that the statistical mechanics should satisfy all
requirements of the equilibrium thermodynamics only in the
thermodynamic limit. Based on a particular example of the ideal
gas we obviously proved the fulfillment of the principle of
additivity and the zero law of thermodynamics for the Tsallis
statistics in the thermodynamic limit. It was shown that all
functions of state of the system are the homogeneous functions of
the first degree, extensive, or the homogeneous functions of the
zero degree, intensive. In particular, the temperature is an
intensive variable and thus provides implementation of the zero
law of thermodynamics. It should be marked that for the finite
values of the number of particles $N$ and the parameter $z$ of the
system both the Tsallis entropy and the Gibbs entropy are
nonextensive functions of state, while in the thermodynamic limit
they become extensive variables. The homogeneous properties of the
functions of state allow us to find the Euler theorem and the
Gibbs-Duhem relation. After applying the thermodynamic limit the
expressions of the Gibbs statistics are obtained by the limiting
procedure $\tilde{z}\to\pm\infty$. The one-particle distribution
function in the thermodynamic limit leads to the Maxwell-Boltzmann
distribution function with the effective mass of particles
$m_{eff}$. This distribution allows us to find the physical
interpretation for the variable of state $\tilde{z}$ as the order
parameter of the interacting system of quasiparticles with mass
$m_{eff}$ and the physical interpretation for the temperature $T$
as the average kinetic energy of quasiparticles. The numerical
example for the ideal gas of quarks with two flavor and three
color degrees of freedom in the nonrelativistic approximation
shows that the dense system of quarks in the dependence of values
of the parameter $\tilde{z}$ can pass from the strong coupled
state to the repulsive state of quarks. In the framework of the
ideal gas of identical particles the equivalence of the canonical
and microcanonical ensembles in the thermodynamic limit is proved.
This property is the key to identifying the self-consistency of
the statistical mechanics and thermodynamics.

{\bf Acknowledgments:} This work has been supported by the
MTA-JINR Grant. We acknowledge valuable remarks and fruitful
discussions with T.S.~Bir\'{o}, R.~Botet, K.K.~Gudima,
M.~P{\l}oszajczak, V.D.~Toneev, and P.~Van.

\appendix
\section{Phase distribution function}
Here, to derive the distribution function in the canonical
ensemble we use the Jaynes principle~\cite{Jaynes}. In this
respect, the Lagrange function can be written as
\begin{equation}\label{a1}
    \Phi[\varrho']=\frac{S[\varrho']}{k} -
    \alpha \left( \int \varrho' \ d\Gamma -1\right)-\beta
    \left( \int \varrho' H \ d\Gamma-\langle H\rangle \right),
\end{equation}
where $\varrho'$ is the probing distribution function. After
maximizing the Lagrange function (\ref{a1}), $\left.
\delta\Phi\right|_{\varrho'=\varrho}=0$, and using Eq.~(\ref{1})
to eliminate the parameter $\alpha$, we arrive at the following
expression for the equilibrium phase distribution function:
\begin{equation}\label{a2}
\varrho=\left[
1+(q-1)\frac{\beta}{q}(\Lambda-H)\right]^{\frac{1}{q-1}},
\end{equation}
where $\Lambda=\langle H\rangle-qS/k\beta$. Differentiating the
function $\Lambda$ and Eq.~(\ref{1}) with respect to $\beta$, and
using the distribution function (\ref{a2}), one finds
\begin{equation}\label{a3}
    \frac{\partial S}{\partial \beta}=k\beta \frac{\partial \langle H\rangle}{\partial
    \beta}.
\end{equation}
The parameter $\beta$ can be related to the temperature
\begin{equation}\label{a4}
    \frac{1}{T}\equiv \frac{\partial S}{\partial E}=
    \frac{\partial S/\partial \beta}{\partial \langle H\rangle/\partial
    \beta}= k\beta, \;\;\;\;\;\;\; \beta=\frac{1}{kT}.
\end{equation}
Then the distribution function (\ref{a2}) takes the form
\begin{equation}\label{a5}
\varrho=\left[
1+(q-1)\frac{\Lambda-H}{kTq}\right]^{\frac{1}{q-1}}=
\left[1+\frac{1}{z+1}\frac{\Lambda-H}{kT}\right]^{z},
\end{equation}
where $\Lambda$ is determined from Eq.~(\ref{2})
\begin{equation}\label{a6}
    \int \left[1+(q-1)\frac{\Lambda-H}{kTq}\right]^{\frac{1}{q-1}}
    d\Gamma=1.
\end{equation}
Note that Eqs.~(\ref{a5}) and (\ref{a6}) are identical with
(\ref{10}) and (\ref{11}). So the form of the distribution
function in terms of the variables of state is independent
of the method of derivation.

Let us show that the distribution function expressed through the
variables of state ($T,V,z,N$) in~\cite{Tsal88} is equivalent to
Eq.~(\ref{a5}). So in~\cite{Tsal88} the Lagrange function was
written as
\begin{equation}\label{a7}
 \Phi[\varrho']=\frac{S[\varrho']}{k} +
    \alpha \int \varrho' \ d\Gamma-\alpha\beta (q-1)
    \int \varrho' H \ d\Gamma.
\end{equation}
After maximizing (\ref{a7}) we obtain
\begin{eqnarray}\label{a8}
 \varrho &=& \frac{1}{Z} \ [1-\beta (q-1)H]^{\frac{1}{q-1}},  \\
  \label{a9}
Z &=&  \int \left[1-\beta (q-1)H\right]^{\frac{1}{q-1}}
    d\Gamma.
\end{eqnarray}
To express the Lagrange parameter $\beta$ through the variables of
state $T,V,z,N$, we use the method described in~\cite{Parv1}.
Finally, for the Lagrange parameter $\beta$ we get
\begin{equation}\label{a10}
    \beta = \frac{Z^{q-1}}{kTq}.
\end{equation}
Substituting Eq.~(\ref{a10}) into Eqs.(\ref{a8}) and (\ref{a9})
and introducing the new function $\Lambda$ in the following form:
\begin{equation}\label{a11}
    Z^{1-q}\equiv 1+(q-1)\frac{\Lambda}{kTq},
\end{equation}
we obtain the phase distribution function (\ref{a5}) with the
normalization condition (\ref{a6}). So the form of the Lagrange
function does not disturb the distribution function in terms of
the variables of state.

\section{The finite perfect gas}

Following the arguments given in Ref.~\cite{Parv1} we easily
derive the norm function $\Lambda$ from Eq.~(\ref{11}) in the case
of $z<-1$:
\begin{equation}\label{1d}
    1+\frac{1}{z+1}\frac{\Lambda}{kT}=\left[Z_{G}\ \frac{\Gamma(-z-\frac{3}{2}N)}
    {(-z-1)^{-\frac{3}{2}N}\Gamma(-z)}\right]^{-\frac{1}{z+\frac{3}{2}N}}\equiv
    B(T,V,z,N), \;\;\;\;\;\;\;\;\;\; z<-1,
\end{equation}
where $Z_{G}=((gV)^{N}/N!)(mkT/2\pi\hbar^{2})^{3N/2}$ is the
partition function of the conventional ideal gas of the Boltzmann-Gibbs
statistics~\cite{Huang,Parv3} and $-z-\frac{3}{2}N>0$. In the case
of $z>0$, we obtain
\begin{equation}\label{2d}
1+\frac{1}{z+1}\frac{\Lambda}{kT}=\left[Z_{G}\
\frac{(z+1)^{\frac{3}{2}N}\Gamma(z+1)}{\Gamma(z+1+\frac{3}{2}N)}
    \right]^{-\frac{1}{z+\frac{3}{2}N}}\equiv
    B(T,V,z,N), \;\;\;\;\;\;\;\;\; z>0,
\end{equation}
where the new function $B$ is introduced for convenience. In order
to determine the energy of system, we insert the Hamilton function
$A(x,p)=H(x,p)$ into Eq.(\ref{12}) and after performing
integration, we obtain
\begin{equation}\label{3d}
    \langle H\rangle=\frac{3}{2}kTN \ \frac{B}{1+\frac{1}{z+1}\frac{3}{2}N},
\end{equation}
where the function $B$ is determined from Eqs.~(\ref{1d}) and
(\ref{2d}). The norm function $\Lambda$ and the energy $\langle
H\rangle$ allow us to calculate the thermodynamic potential of the
canonical ensemble, the free energy $F$. Substituting
Eqs.(\ref{1d}), (\ref{2d}) and (\ref{3d}) into Eq.(\ref{14}), we
find
\begin{equation}\label{4d}
    F=-kTz\left[1- \frac{\left(1+\frac{1}{z}\frac{3}{2}N\right)B}
    {1+\frac{1}{z+1}\frac{3}{2}N}\right].
\end{equation}
Then, the entropy of the system can be easily obtained from
Eqs.(\ref{13}) or (\ref{14}):
\begin{equation}\label{5d}
    S=kz\left[1-\frac{B}{1+\frac{1}{z+1}\frac{3}{2}N}\right].
\end{equation}

Let us calculate the pressure $p$, the chemical potential $\mu$,
and the variable $X$. Taking into account Eqs.~(\ref{7b}) and
(\ref{4d}), one finds
\begin{equation}\label{6d}
    p=\frac{N}{V}kT\ \frac{B}{1+\frac{1}{z+1}\frac{3}{2}N} =
    \frac{2}{3}\ \frac{E}{V}.
\end{equation}
Differentiating Eq.~(\ref{4d}) with respect to $z$ in conformity
with Eq.~(\ref{8b}), we obtain
\begin{equation}\label{7d}
    X=kT\left(1-\frac{B}{1+\frac{1}{z+1}\frac{3}{2}N}\
    \left[1-\frac{(\frac{1}{z+1})^{2}\frac{3}{2}N}{1+\frac{1}{z+1}\frac{3}{2}N}\
    -\ln B +\psi\left(a+\gamma\frac{3}{2}N\right)-\psi(a)\right]\right),
\end{equation}
where $\psi(y)$ is the psi-function which depends on arguments
$a=-z$, $\gamma=-1$ for $z<-1$ and $a=z+1$, $\gamma=1$ for $z>0$.
Taking the derivative of (\ref{4d}) with respect to $N$ in
conformity with Eq.~(\ref{8b}), we get
\begin{eqnarray}\label{8d}\nonumber
    \mu &=& \frac{3}{2}kT \frac{B}{1+\frac{1}{z+1}\frac{3}{2}N}\
\left[\frac{\frac{1}{z+1}}{1+\frac{1}{z+1}\frac{3}{2}N}\
-\ln(\gamma(z+1)B)+\psi\left(a+\gamma\frac{3}{2}N\right)\right] - \\
&-& kT\frac{B}{1+\frac{1}{z+1}\frac{3}{2}N}\
\left(\ln\left[V\left(\frac{mkT}{2\pi\hbar^{2}}\right)^{3/2}\right]
-\psi(N+1)\right).
\end{eqnarray}

The $N$-particle distribution function of the classical ideal gas
in the canonical ensemble for the Tsallis statistics can be written as
\begin{equation}\label{9d}
    f(\vec{p}_{1},\ldots,\vec{p}_{N}) =\frac{(gV)^{N}}{N!h^{3N}}
    \left[1+\frac{1}{z+1}
    \frac{\Lambda-\sum_{i=1}^{N}\frac{\vec{p}_{i}^{2}}{2m}}{kT}\right]^{z},
\end{equation}
which is normalized to unity
\begin{equation}\label{10d}
    \int d^{3}p_{1}\cdots d^{3}p_{N}
    f(\vec{p}_{1},\ldots,\vec{p}_{N})=1.
\end{equation}
Then the reduced single-particle distribution function can be
easily obtained by directly performing the integral over the
momenta $(\vec{p}_{2},\ldots,\vec{p}_{N})$:
\begin{equation}\label{11d}
    f(\vec{p})=\left[\frac{\Gamma(a+\gamma\frac{3}{2}N)}
    {\Gamma(a+\gamma\frac{3}{2}(N-1))}\right]^{\gamma}
    \left(\frac{\gamma}{(z+1)2\pi m kT B}\right)^{3/2}
    \left[1-\frac{1}{z+1}\frac{\vec{p}^{2}}{2mkTB}\right]^{z+\frac{3}{2}(N-1)}.
\end{equation}

\end{document}